\documentclass[a4paper,fleqn,usenatbib,useAMS]{mnras}

\usepackage{graphicx}	
\usepackage{amsmath}	
\usepackage{amssymb}	
\usepackage{multicol}        
\usepackage{bm}		
\usepackage{pdflscape}	

\usepackage[T1]{fontenc}
\usepackage{ae,aecompl}

\usepackage{newtxtext,newtxmath}

\title[Systematic Regularity of Solar Coronal Rotation]{Systematic Regularity of Solar Coronal Rotation During the Time Interval 1939-2019}

\author[L. H. Deng et al.]{
L. H. Deng,$^{1,3,4,5}$
X. J. Zhang,$^{3,7}$
H. Deng,$^{2,6}$
Y. Mei,$^{2}$
and F. Wang$^{2,3,6}$\thanks{E-mail: fengwang@gzhu.edu.cn (FW)}
\\
$^{1}$Chongqing University of Arts and Sciences, Chongqing~402160, P.R.~China\\
$^{2}$Center For Astrophysics, Guangzhou University, Guangzhou~510006, P.R.~China\\
$^{3}$Yunnan Observatories, Chinese Academy of Sciences, Kunming~650216, P.R.~China\\
$^{4}$College of Science, China Three Gorges University, Yichang~443002,P. R. China\\
$^{5}$Key Laboratory of Geospace Environment, Chinese Academy of Sciences, University of Science \& Technology of China, Hefei~230026, P.R.~China\\
$^{6}$CAS Key Laboratory of Solar Activity, National Astronomical Observatories, Beijing~100012, P.R.~China\\
$^{7}$University of Chinese Academy of Sciences, Beijing~100049, P.R.~China}

\date{Accepted 2019 November 05. Received 2019 October 14; in original form 2019 July 24}

\pubyear{2019}

\begin{document}
\label{firstpage}
\pagerange{\pageref{firstpage}--\pageref{lastpage}}
\maketitle

\begin{abstract}
Temporal variation of the solar coronal rotation appears to be very complex and its relevances to the eleven-year solar activity cycle are still unclear. Using the modified coronal index for the time interval from 1939 January 1 to 2019 May 31, the systematic regularities of the solar coronal rotation are investigated. Our main findings are as follows: (1) from a global point of view, the synodic coronal rotation period with a value of 27.5 days is the only significant period at the periodic scales shorter than 64 days; (2) the coronal rotation period exhibit an obviously decreasing trend during the considered time interval, implying the solar corona accelerates its global rotation rate in the long run; (3) there exist significant periods of 3.25, 6.13, 9.53, and 11.13 years in the period length of the coronal rotation, providing an evidence that the coronal rotation should be connected with the quasi-biennial oscillation, the eleven-year solar cycle, and the 22-year Hale cycle (or the magnetic activity reversal); and (4) the phase relationship between the coronal rotation period and the solar magnetic activity is not only time-dependent but also frequency-dependent. For a small range around the 11- year cycle band, there is a systematic trend in the phase, and the small mismatch in this band brings out the phase to drift. The possible mechanism for the above analysis results is discussed.
\end{abstract}

\begin{keywords}
Sun: corona -- Sun: activity -- Sun: rotation
\end{keywords}




\section{Introduction}

%
%

Solar rotation is one of the essential properties of the Sun as a star, but it is still to be understood clearly from the existing large volumes of observational data \citep{1985SoPh..100..141S,2000SoPh..191...47B,2014AJ....148...12X}. Generally speaking, traditional measurements of solar differential rotation in the surface atmospheric layers could be carried out mainly by using three methods: magnetic tracers, spectroscopic observations, and flux modulations \citep{1974ARA&A..12...47G,1984ARA&A..22..131H,2001ApJ...548L..87V}. Besides, the understanding of the interior rotation of the Sun varying with latitude and radius is also very important, and is usually determined by using the techniques developed through helioseismology \citep{1985Natur.317..591B,2003ARA&A..41..599T,2009LRSP....6....1H}. However, each of these methods has its own difficulties and limitations. That is to say, the high-precision determination of solar rotation periodicity is not an easy work as it depends on both the types of observational data and the analysis approaches.

Previous studies showed that there is a general consensus on the magnitude and form of the interior rotation as inferred from helioseismology \citep{1998MNRAS.298..543A}, the surface differential rotation rate as derived from sunspots \citep{2013SoPh..287..197J}, Doppler velocity measurements \citep{1990ApJ...351..309S}, as well as the magnetic activity features \citep{1993SoPh..145....1K,2016AJ....151...76X}, but there is no such consensus regarding the rotation rule for the features in the solar corona, as pointed out by \cite{2013ApJ...763..137H}. The main reasons why the coronal rotation is relatively less determined are as follows: i) there are virtually no obvious tracers, ii) the corona is the optically thin region, and iii) it is difficult to measure the coronal magnetic field \citep{1998SoPh..181..351V,2018MNRAS.475.3117B}. At present, it is widely accepted that the rotation of the solar corona could provide additional information on the construction of solar dynamo models, because it reflects the rotation characteristics in the sub-photospheric layers of the Sun \citep{2013IAUS..294..399K,2018AJ....156..152X,2018SoPh..293..128B}. 

Lots of researchers have analyzed the coronal line emission at \normalsize{5303} \AA~ (Fe \tiny{XIV} \normalsize{green} line) to investigate the rotation characteristics of the solar corona \citep{1989ApJ...336..454S,1994SoPh..152..161R,1999SSRv...87..211I,2003SoPh..213...23A,2006AdSpR..38..906B,2010NewA...15..135B,2017MNRAS.466.4535B}. To reveal the rotational behavior of the solar corona in the Fe \tiny{XIV} \normalsize{green} line, \cite{1989ApJ...336..454S} analyzed the observational data made with the Sacramento Peak 40 cm coronagraph between 1973 and 1985. They found that the synodic rotation period is 27.52$\pm$0.42 days for the latitudes from 30N to 30S, and the corona rotates more rigidly than the features in the lower atmosphere. Using the homogeneous data set of coronal green line intensities over the period 1964-1990, \cite{1994SoPh..152..161R} calculated the averaged synodic rotation period as a function of latitude and time, and found that the coronal rotation period for the whole range of latitudes and for the latitudinal band $\pm30^\circ$ are 28.18$\pm$0.12 days and 27.65$\pm$0.13 days, respectively. Based on the data observed by the SOHO/LASCO C1 coronagraph during the solar minimum from April 1996 to March 1997, \cite{1999SSRv...87..211I} studied the periodicity and recurrence of Fe \tiny{XIV} \normalsize{emission} structures with heliospheric latitude and distance above the Sun's surface, they found no significant deviation from a rigidly rotating Fe \tiny{XIV} \normalsize{corona} with latitude or with distance from the Sun even on the small scales. Using the database of the coronal green line brightness (CGLB) covering six solar cycles (1939-2001), \cite{2006AdSpR..38..906B} found that the period of coronal rotation increases from 27 days at the solar equator to a little more than 29 days within the latitudes $\pm40^\circ$. Moreover, they found that, for the higher latitudes, the rotation period changes very slowly and its value of about 29 days remains up to the polar regions. That is to say, the total rotation of the solar corona could be described as a superposition of two rotation modes: the fast mode near the equator is around 27 days, and the slow mode exceeds 30 days \citep{2010NewA...15..135B}. Using the same but longer database, \cite{2017MNRAS.466.4535B} investigated the time variation of the differential rotation parameters of the solar corona, and found that the equatorial rotation rate of the corona increases in the epochs of minimum between the even and odd cycles and arrives at its minimum values between the odd and even cycles. They suggested that the 22-year cycle of differential rotation pattern could be used to explain the effect by the Gnevyshev-Ohl rule.

Meanwhile, solar radio fluxes at different frequencies are also used to determine the temporal variation of solar coronal rotation \citep{1997EM&P...76..141V,1998SoPh..181..351V,2001ApJ...548L..87V,2009MNRAS.400L..34C,2011MNRAS.413L..29V,2011MNRAS.414.3158C,2012MNRAS.423.3584L,2017ApJ...841...42X,2017SoPh..292...55B,2018MNRAS.475.3117B}. By using the disk-integrated simultaneous measurements of solar radio flux at the frequency range of 275-2800 MHz,  \cite{2001ApJ...548L..87V} found that the coronal rotation period strongly depends on the heights in the corona, i.e. the sidereal coronal rotation period increases with increasing frequency. However, by using the same data sets, \cite{2017SoPh..292...55B} found that the solar corona rotates more slowly at higher altitudes, which contradicts the conclusion given by \cite{2001ApJ...548L..87V}. By studying the coronal rotation period as a function of latitude between $\pm60^\circ$ for the period from 1999 to 2001, \cite{2009MNRAS.400L..34C} found that the solar corona rotates less differentially than the photosphere and chromosphere. The similar but abundant results were confirmed by \cite{2010MNRAS.407.1108C} who studied the differential rotation of soft X-ray corona derived from the solar full disk images of soft X-ray telescope on board the $Yohkoh$ observatory. Using the 10.7 cm solar radio emission data for the time interval 1947-2009, \cite{2011MNRAS.414.3158C} found that sidereal coronal rotation period varies from 19.0 days to 29.5 days with an average of 24.3 days, and a 22-year component (which might be related to the 22-year Hale magnetic cycle or magnetic field reversal) exists in the long-term rotation variation. Based on their analysis results,  \cite{2012MNRAS.423.3584L} investigated the long-term variation of the coronal rotation of radio emission, and found a weak decreasing trend in the coronal rotation period. They found that there is no 11-year Schwabe cycle of statistical significance for the secular variation of rotation period length, and the solar coronal rotation does not exhibit any systematic pattern. By applying the cross-correlation analysis and wavelet transformation, \cite{2017ApJ...841...42X} studied the daily and monthly radio flux at 2800 MHz from 1947-2014. They found that the coronal rotation period varies with the solar cycle phase, and the rotation period is relatively longer around the minimum year of a solar cycle. Interestingly, they found the coronal rotation variation is related to the 11-year Schwabe cycle, which is disagreement with the result obtained by \cite{2011MNRAS.414.3158C} and \cite{2012MNRAS.423.3584L}. In addition, the variability of fractal dimension of solar radio flux at different frequencies is also studied by \cite{1997EM&P...76..141V} and \cite{2018MNRAS.475.3117B}, and the main conclusion is that the fractal dimension increases with increasing frequency, i.e. randomness increases towards the inner corona.

Clearly, the long-term variation of solar coronal rotation and its relationship with the eleven-year solar activity cycle are not well understood, more studies on these topics are thus required and needed. With the hope to add further information on the temporal variation and the underlying processes of solar coronal rotation, we focus on modified coronal index for nearly eight solar cycles, introduced and provided by the Slovak Central Observatory in Hurbanovo, through several time series analysis approaches. A brief introduction of the observational data used in this paper is shown in the next Section. In Section 3 the statistical analysis results for coronal rotation determination are presented. And finally the conclusions and discussions are given in the last Section.

\section{Observational Data}

\subsection{Modified Coronal Index}
\label{sec:data} 

The coronal green line (Fe \tiny{XIV} \normalsize{5303} \AA) is the most typical emission line from the inner corona of the Sun, and the regular observations of this prominent line have been processed since the year of 1939 \citep{2005JGRA..110.8106R}. By searching a suitable measurement from a Sun-observing space-based probe, \cite{2010SoPh..263...43L} introduced the modified coronal index (MCI) to replace the original coronal index, which was first proposed by \cite{1975BAICz..26..367R}.

As pointed out by \cite{2010SoPh..263...43L}, MCI represents the physical measure of solar cycle signature in the outer atmosphere and its advantage is the longest time sequence. This indicator describes the average irradiance of green line emitted in one steradian towards the Earth, and the daily values of MCI from 1939 January 1 to 2019 May 31 are shown in Figure 1. Here, the time series of MCI is downloaded from the website of the Slovak Central Observatory in Hurbanovo\footnote{http://www.suh.sk/obs/vysl/MCI.htm}, and it is expressed in ($10^{-6}\rm W/m^2$) units.

\begin{figure}
   \includegraphics[width=\columnwidth]{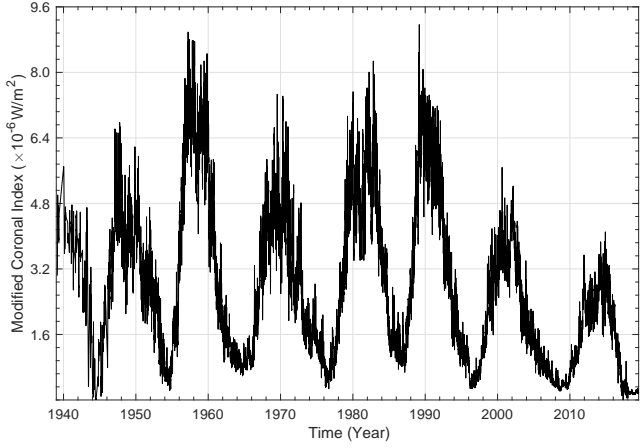}
    \caption{The daily values of modified coronal index (unit: $10^{-6}\rm W/m^2$) for the time interval from 1939 January 1 to 2019 May 31.}
    \label{fig:figure1}
\end{figure}

\begin{figure}
    \includegraphics[width=\columnwidth]{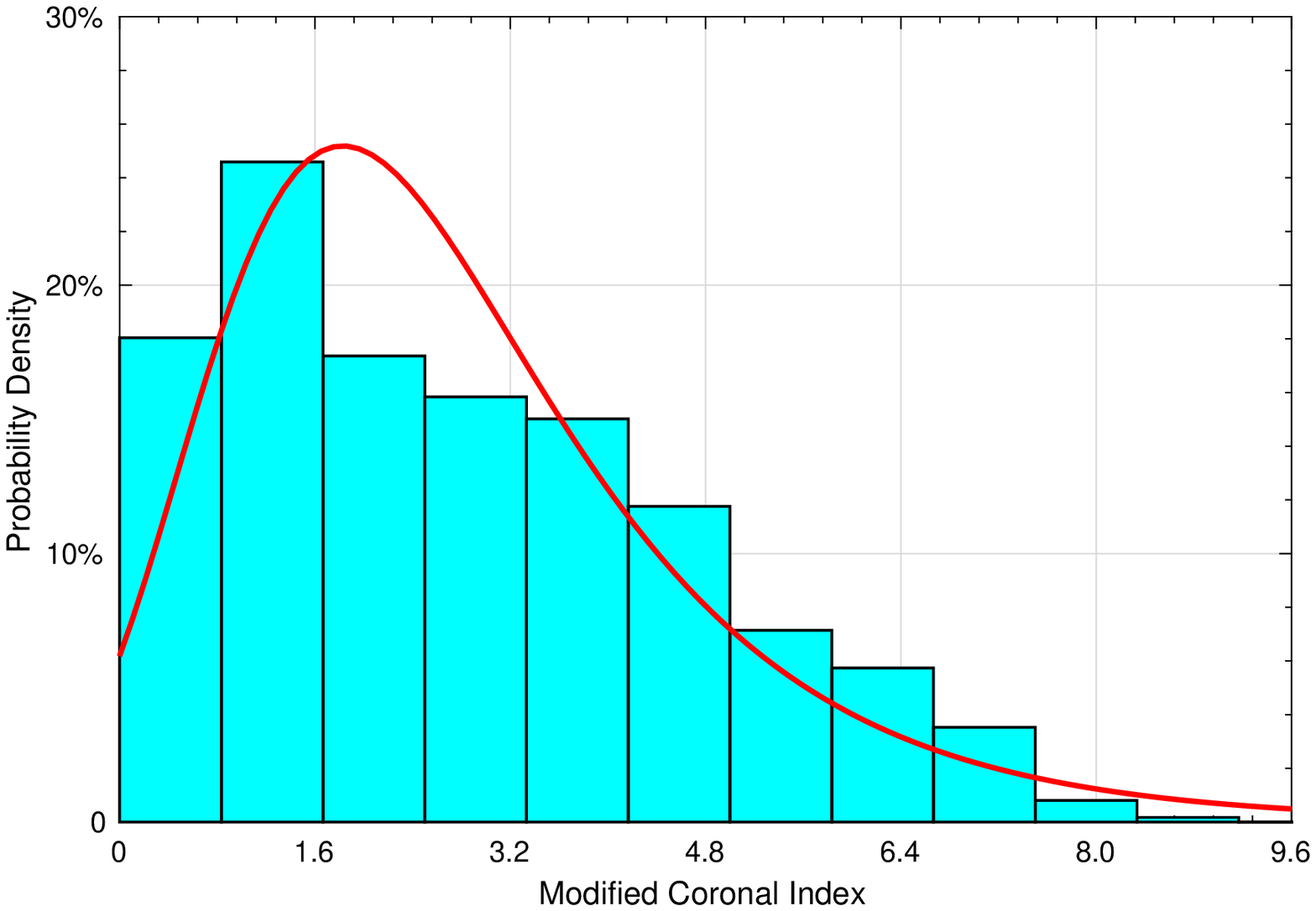}
    \caption{The histogram distribution of the probability density for the MCI measurements (the cyan blocks). The probability distribution can be described by the GEV distribution (the red line).}
    \label{fig:figure2}
\end{figure}

\begin{figure}
    \includegraphics[width=\columnwidth]{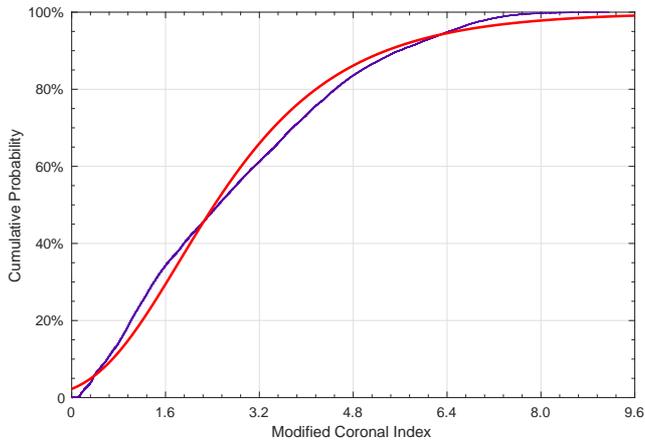}
    \caption{The cumulative probability distribution of the MCI measurements (the blue line) and the fitting values (the red line).}
    \label{fig:example_figure}
\end{figure}

\begin{figure}
   \includegraphics[width=\columnwidth]{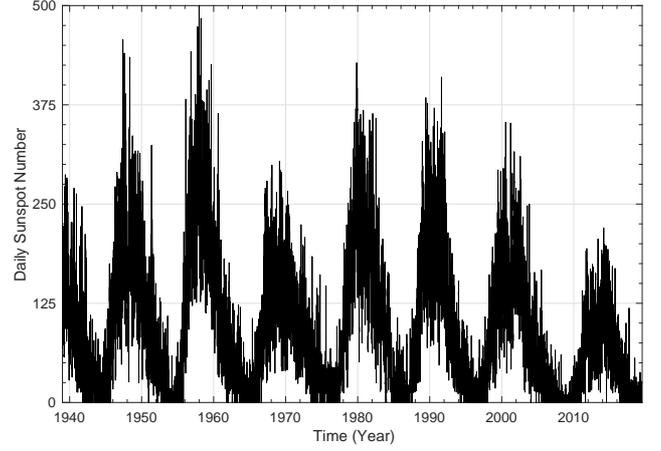}
    \caption{The daily values of sunspot number for the time interval from 1939 January 1 to 2019 May 31.}
    \label{fig:figure1}
\end{figure}

\begin{figure}
    \includegraphics[width=\columnwidth]{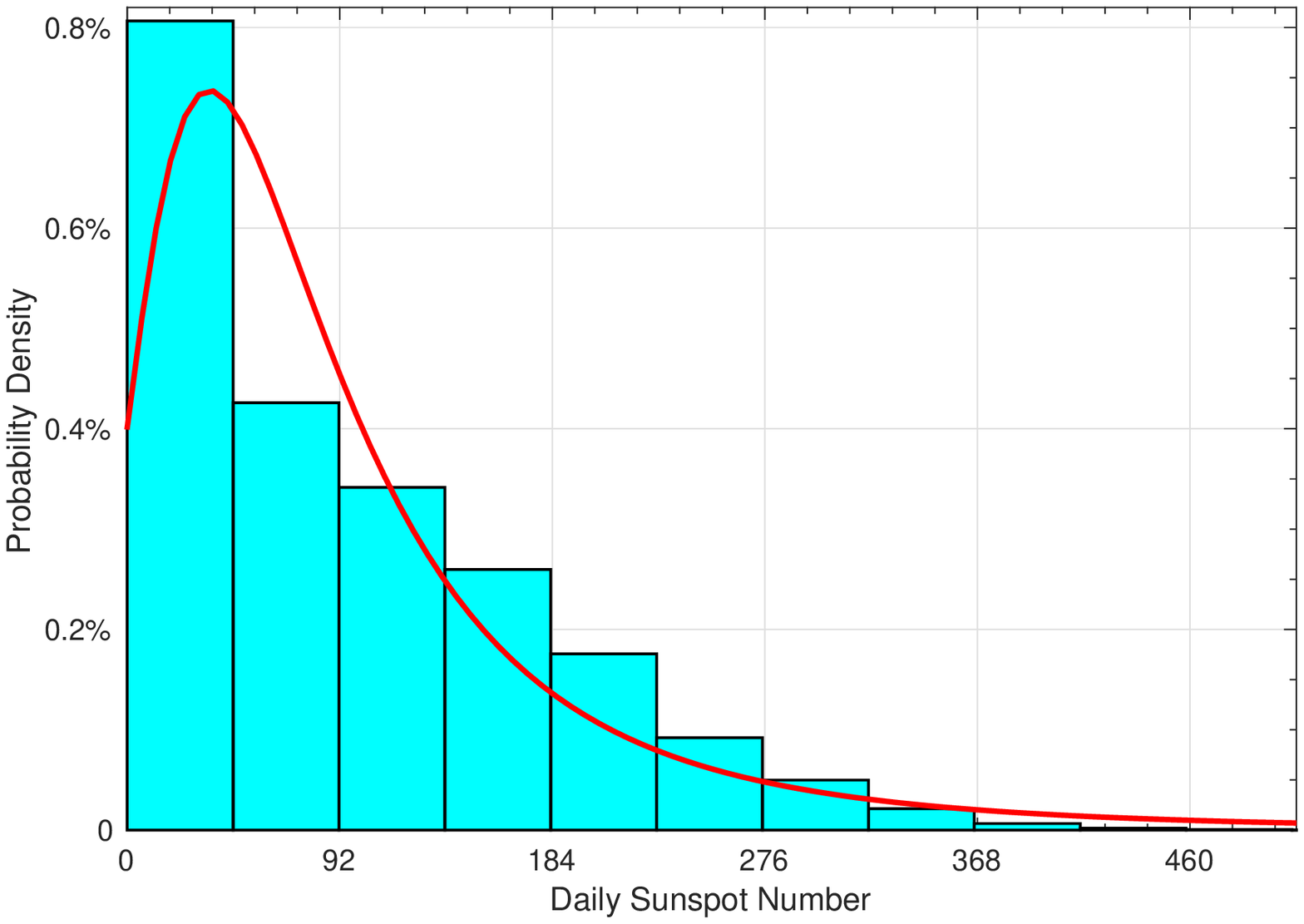}
    \caption{The histogram distribution of the probability density for the DSN values (the cyan blocks). The probability distribution can be described by the GEV distribution (the red line).}
    \label{fig:figure2}
\end{figure}

\begin{figure}
    \includegraphics[width=\columnwidth]{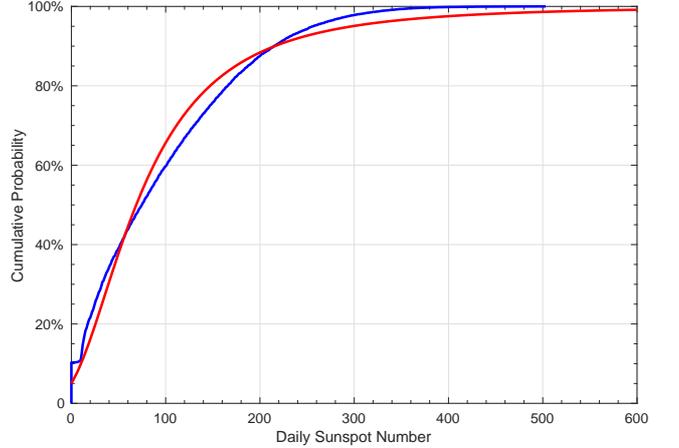}
    \caption{The cumulative probability distribution of the DSN values (the blue line) and the fitting values (the red line).}
    \label{fig:example_figure}
\end{figure}

\subsection{Probability Distribution}
\label{sec:data} 

Since we are working with the truncated sets, the truncated distributions on the fit --- building the data set from each probability distribution function (PDF) and cumulative distribution function (CDF) are used. To look for the probability distribution of the MCI values, they are divided into twelve bands with each band spanning 0.83. The numbers of MCI values within each band are counted, and then they are divided by their total numbers to describe the probability density.  

Figure 2 displays the histogram distribution of the probability density of the MCI values. Here, the ordinate is the ratios defining the percentage probability of the MCI values for each band.

In order to find the best fit distribution of the data set, the distributions of the exponential, logistic, normal, and generalized extreme value (GEV) are examined. We found that, among the four distribution functions, the probability distribution of the MCI values can be described by the GEV distribution that models block maxima of the variables \citep{2007A&A...472..293A}.

 The red line shown in Figure 2 is the GEV distribution, and the most probable value of the MCI measurements is found to be 1.93. The cumulative probabilities of the MCI measurements (the blue line) and the fitting values (the red line) are shown in Figure 3, and one can see that the two curves match quite well.
 
 \subsection{Statistic Test of the Distribution}

Ultimately, we need to test the relative performance of the GEV distribution that is used to fit the data. To quantify the relative performance, the Kolmogorov-Smirnov (K-S) statistic test, which corresponds to the biggest difference between the observed and the empirical CDFs, is used.

\begin{equation}
\mathrm{KS} = \max \mid \mathrm{CDF}_{\mathrm{obs}}(x) -\mathrm{CDF}_{\mathrm{emp}}(x) \mid
\end{equation}

where $x_{\mathrm{trunc}} \leqslant x \leqslant \infty$, and $x_{\mathrm{trunc}}$ is the limit value below which data is not used in the fit.

Unfortunately, the GEV distribution does not pass the K-S test (in which the null hypothesis assumes that the MCI data is drawn by the GEV distribution), because the value of the KS (here is 0.0397) is larger than the Table of the percentage points of Kolmogorov statistics (the value for the 29371 data points is 0.0071 at the 95\% confidence level) that was provided by \cite{miller1956table}.  We also examined the other three distributions (exponential, logistic, and normal distribution), however, all of them did not pass the K-S test. That is, the distribution of the MCI data is not the GEV distribution, although the GEV distribution is the best one among the four distributions.

 \subsection{Comparison with the Sunspot Number Data}
 
 \cite{2002SoPh..206..401M} analyzed the correlation of the coronal index and other solar indices for the time period 1965-1997. They found that the coronal index could be used as a representative index of solar activity in order to be correlated with different periodic solar-terrestrial phenomena useful for space weather studies. To confirm the features that cause the coronal modulation are rooted in surface magnetic structures of the Sun, here we directly compared the distribution of the MCI data with the sunspot indicator.

The time series of the daily sunspot number (DSN) used in this work is freely downloaded from the World Data Center (WDC) --- Sunspot Index and Long-term Solar Observations (SILSO), Royal Observatory of Belgium, Brussels\footnote{http://www.sidc.be/silso/datafiles}. Since 2015 July 1, the original version of the DSN data has been replaced by a new entirely revised data set (the so-called version 2.0). For details, please refer to \cite{2014SSRv..186...35C} and \cite{2016SoPh..291.2629C}. Here, the time period from 1939 January 1 to 2019 May 31, the common time interval to the MCI data, is extracted. Figure 4 displays the daily values of sunspot number during the past 80 years (1939-2019).

Similar to the MCI data, we also divided the DSN time series into twelve bands, and then calculated the PDF for each band. Four distributions (exponential, logistic, normal, GEV) are used to find the best fit distribution of the DSN data, and the results are shown in Figures 5 and 6. The results indicate that the probability distribution of the DSN can be described by the GEV distribution, but it does not pass the K-S test. That is to say, the distribution of the DSN data is not the GEV distribution, although the GEV distribution is the best one among the four distributions.

From the above analyses, we can arrive at a conclusion that both the MCI and the DSN data show the similar distribution behavior, but they are not the GEV distribution.

However, it should be pointed out that the extreme value theory can be used to study the solar extreme events. For example, \cite{2017ApJ...839...98A,2018ApJ...853...80A,2019SoPh..294...67A} employed this technique to study the temporal variability of the sunspot number and the solar radio flux, and found that there is an upper bound for the distribution of the peak values of the photospheric and coronal indicators. Therefore, the coronal magnetic activity exhibits the similar kind of behavior as the sunspot activity, implying that the features that cause the coronal modulation are rooted in surface magnetic structures.

\section{Analysis Results}

\subsection{Period Length of Coronal Rotation}

To determine the coronal rotation period of MCI varying with the time, the codes of wavelet transform analysis provided by \cite{1998BAMS...79...61T}\footnote{http://atoc.colorado.edu/research/wavelets/} is utilized. Before the method is applied, the average values of MCI are normalized first, which is a procedure that subtracts the average value and then divides by the variance of the data set. In our analysis, the statistical significance test is carried out by assuming that the red noise exists in the given time series, which is also described in the above paper. After the normalized procedure, the resulting local power spectra of MCI are displayed in Figure 7. As we focus on studying the temporal variation of solar coronal rotation (around 27 days), the horizontal dashed black lines, representing the periods of 26 days and 38 days, are used to show the relative higher power belts.

\begin{figure}
    \includegraphics[width=\columnwidth]{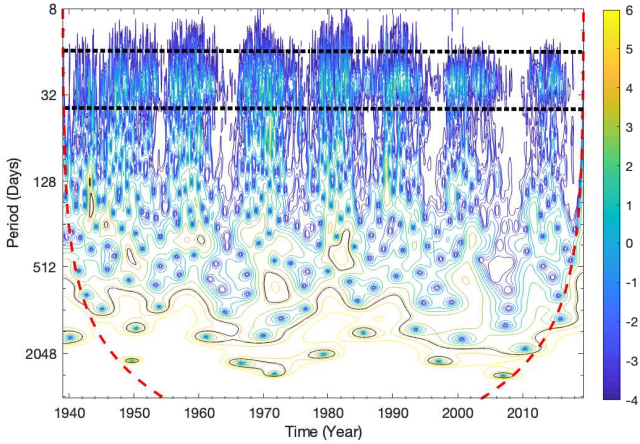}
    \caption{Local wavelet power spectrum of the daily MCI, and the horizontal dashed black lines are the periods of 26 and 38 days. The dashed red line is the cone of influence (COI) where the edge affects might distort the pattern.}
    \label{fig:example_figure}
\end{figure}

As shown in Figure 7, it can be seen that the relative higher power belts appear around 27 days at the maximum times of solar cycles, as the signal-to-noise is relative higher when the MCI itself is higher. For the minimum times of solar cycles, the solar corona is also rotating. However, the signal-to-noise is relative lower at the periods of lower activity, so the rotational behavior is difficult to seen from the local wavelet power spectra. According to Figure 7, the global wavelet power spectra of daily MCI with the periodic scale up to 128 days are calculated and shown in Figure 8. From this figure one can clearly see that the coronal rotation period with a value of 27.5 days is the only one peak at the periodic scales shorter than 64 days, which is statistically significant above the 95\% confidence level (dashed red line). It is noteworthy that the periodicity obtained here is the synodic coronal rotation period ($\rm P_{synodic}$), so the sidereal coronal rotation period ($\rm P_{sidereal}$) is determined to be 25.57 days. The approximate conversion relation between them could be described by the following expression \citep{1995SoPh..159..393R,1996SoPh..168..211W,2014SoPh..289.1471S}:

\begin{equation}
\rm P_{sidereal} = \frac {365.26 \times P_{synodic}} {365.26+P_{synodic}}
\end{equation}

The synodic coronal rotation period (27.5 days) obtained by us is good agreement with the results reported by previous researchers, whose studies focused on the Fe \tiny{XIV} \normalsize{green} line for the period 1973-1985 \cite[27.52$\pm$0.42 days]{1989ApJ...336..454S}, the CGLB over the period 1964-1990 \cite[27.65$\pm$0.13 days]{1994SoPh..152..161R}, and the Fe \tiny{XIV} \normalsize{emission} structures observed by SOHO/LASCO C1 coronagraph \cite[27.5$\pm$1 days]{1999SSRv...87..211I}. However, the obtained sidereal coronal rotation period in this work (25.57 days) is slightly greater than the result (24.3$\pm$0.2 days) given by \cite{2011MNRAS.414.3158C} who studied the daily radio emission at 2.8 GHz between 1947 and 2009.

\begin{figure}
    \includegraphics[width=\columnwidth]{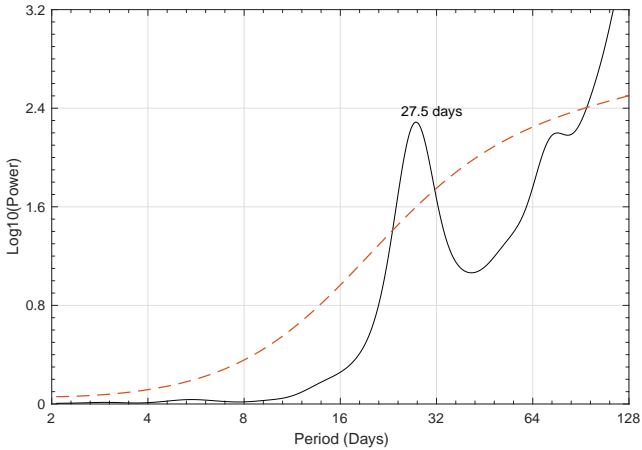}
    \caption{Global wavelet power spectrum of the daily MCI with the periodic scale up to 128 days, and the dashed red line is the 95\% confidence level.}
    \label{fig:example_figure}
\end{figure}

\begin{figure}
    \includegraphics[width=\columnwidth]{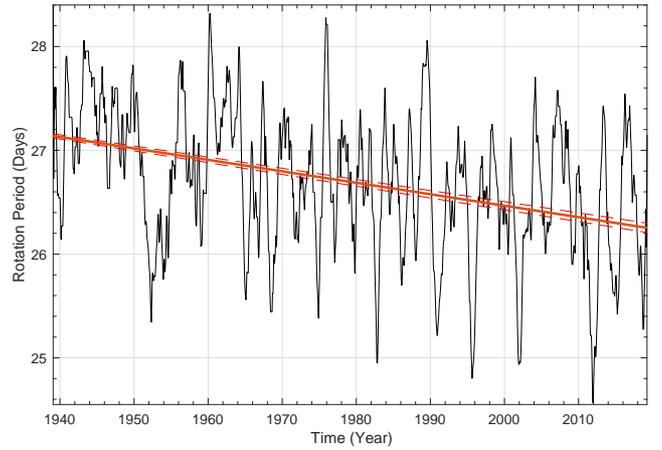}
    \caption{The temporal variation of the period length of the coronal rotation (PLCR) during the considered time interval. The bold solid red line shown in this figure is the linear regression line, and the dashed red lines are the 95\% confidence bounds.}
    \label{fig:example_figure}
\end{figure}

\subsection{Temporal Variation of Coronal Rotation}

Since the local wavelet power spectra provide detailed contents in the time-frequency space, the local periodicity or frequency of daily MCI at a particular time point could be estimated. That is to say, in the local wavelet power spectra shown in Figure 7, the rotation period length at a certain time point has the highest power spectrum among the periodic scales of 24-30 days, where are the length range of coronal rotation period located \citep{1984ApJ...283..373H,2011MNRAS.414.3158C}. Based on this conception, the rotation period length of sunspot numbers \citep{2011Ap&SS.331..441L,2012RAA....12..187X}, sunspot areas \citep{2011ApJ...730...49L}, solar mean magnetic field \citep{2017AJ....153..171X}, and 10.7 cm solar radio flux \citep{2017ApJ...841...42X} are studied and estimated. In this sense, the similar procedure could be also applied to obtain the period length of the coronal rotation (PLCR) of daily MCI at each time point during the considered time interval, and the extracted results are shown in Figure 9. The time series of the PLCR is smoothed with a one-year sliding window, and the rotation period shown here is also the synodic coronal rotation period (the average value is 26.69$\pm$0.68 days).

As Figure 9 displays, the PLCR varies from cycle to cycle, and a general trend with decrease in length is clearly seen. To look into the secular trend of the PLCR time series, we perform a linear regression analysis to fit the data set, and the regression line (indicated by the bold solid red line) is also shown in this figure. The dashed red lines indicate the 95\% confidence bounds for the linear regression analysis. The relation between the two parameters can be described as $p(t)$ = -0.0110 ($\pm$0.0003) $*$ $t$ + 27.14 ($\pm$0.02), where $t$ is the time (in year, here the start time is set to 0) and $p(t)$ is the PLCR varying with the $t$. From the linear regression analysis, the PLCR is found to have a general trend, which exhibits a linear decrease from 27.14 ($\pm$0.02) days to 26.25 ($\pm$0.01) days (the period difference is 0.89 ($\pm$0.03) days with a slope of about -0.0110 ($\pm$0.0003) day/year). 

To quantitatively judge the result of fit, the goodness-of-fit with 95\% confidence is used. The parameter of Adjusted R-square (Adj. R$^2$) value is calculated and represents the most useful measure of the success of the statistical model. In practice, if this value is close to zero, the fit result can be considered as a good fit. Another statistical parameter, the RMSE (root-mean-squared error), which is known as the fit standard error and the standard error of the regression, is also used. Similar to the Adj. R$^2$ value, the RMSE value close to zero indicates a good fit. Here, the Adj. R$^2$ value and the RMSE are 0.1416 and 0.6316, respectively. These two parameters are not so large that the fit result can be considered as a good fit. Therefore, at least from a global point of view, the solar corona rotates more and more speedily, i.e. accelerates its rotation rate, during the studied time interval (1939-2019).

It is clear from Figure 9 that the long-term decreasing trend of PLRC time series we obtained is similar to those obtained by \cite{2009A&A...497..835H}, \cite{2011MNRAS.414.3158C},  \cite{2012MNRAS.423.3584L}, and \cite{2017ApJ...841...42X}. Interestingly, all of these studies used the daily 10.7 cm solar radio flux to reveal the rotational behavior of the solar corona. Since both the 10.7 cm radio flux and the Fe \tiny{XIV} \normalsize{green} line emission reflect the global characteristic of the solar corona, the long-term temporal variation of PLCR obtained by these two indicators are similar to each other. A similar inference could be drawn at the photospheric magnetic activity estimated by \cite{2011ApJ...730...49L} who studied the rotational cycle length of daily sunspot areas from 1849 to 2010. They found that the rotation length of the photospheric rotation have a secular trend showing a linear decrease by about 0.47 days. However, the period difference of the PLCR obtained by us (0.89 $\pm$0.03 days) is almost twice as large as that of \cite{2011ApJ...730...49L}. The possible reason for the significant difference of the two works is that the studied time interval differs from each other. The time interval of the sunspot number used in \cite{2011ApJ...730...49L} is from 1849 January 1 to 2010 February 28 (about 162 years), whereas the time interval of the MCI data used in work is only 81 years. By using the similar fit method, a smaller period difference can be seen from longer time series. This reason can be confirmed by pervious works. For example, the time interval of the sunspot number used in \cite{2009A&A...497..835H} is from 1849-2004 (about 156 years), their work gave a value of 0.66 days. That is, the period differences obtained by different authors are not caused by the uncertainties of the rotation period. So from the present work and earlier studies, we infer that sunspot activity, coronal flux, and green corona exhibit the similar global rotation behavior.


\subsection{Periodicity in the PLCR Time Series}

There are many time-frequency analysis methods to search the periodic variation of solar time series. One of such statistical techniques is the auto-correlation analysis, which has successfully been employed for determining the periodicities of solar activity indicators \citep[see][]{1969SoPh...10..135H,1998SoPh..181..351V,2001ApJ...548L..87V,2010MNRAS.407.1108C}. By applying this technique to the PLCR time series, the auto-correlation coefficients are plotted against the phase lags (in days), and the analysis result is shown in Figure 10. We can easily find that, for the phase shifts smaller than 4500 days, there are four prominent peaks when the phase shifts are 3.25, 6.13, 9.53, and 11.13 years. The dashed red line shown in Figure 10 is the 95\% confidence level line, so the four periodicities obtained by us are statistical significant.

The periodicity of 3.25 years is the so-called quasi-biennial oscillation (QBO; \citealt{2014SSRv..186..359B}), and the similar periods of 2-3 years were also found by \cite{1995SoPh..158..173J}, \cite{1999SoPh..184...41J}, \cite{2009SoPh..257...61J}, and \cite{2017ApJ...841...42X}. These studied focused on the temporal variations of solar rotation derived from the sunspot group data, Doppler-velocity measurements, and daily 10.7 cm radio flux. Our analysis results further support the existence of QBO in the periodic variation of solar coronal rotation. Although the physical mechanism of solar QBO has not been fully understood, it is believed to be intrinsic to the internal dynamo process. Moreover, previous studies showed that the source of the QBO of the solar magnetic field should be situated at the base of the solar convection zone. Most recently, \cite{2010ApJ...724L..95Z}, \cite{2015NatCo...6E6491M}, and \cite{2018ApJ...856...32Z} concluded that the solar QBO of magnetic activities should be caused by the magnetic Rossby wave instabilities. Therefore, it is an evidence that the rotational behavior of the solar corona shows the similar temporal variation as the lower atmospheric layers, and both of them might be connected with dynamical process at the base of the solar convection zone.

\begin{figure}
    \includegraphics[width=\columnwidth]{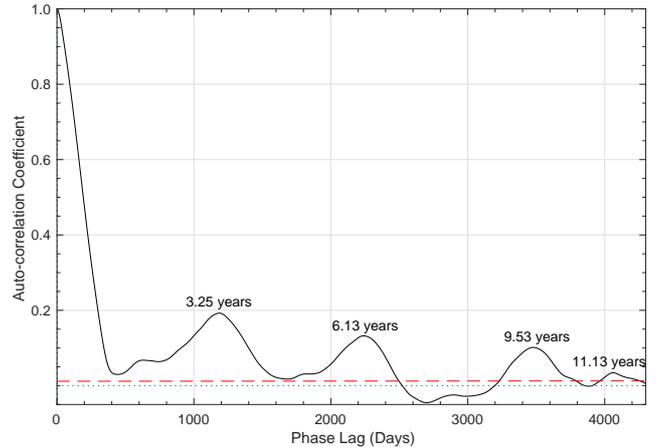}
    \caption{The auto-correlation coefficient of the daily PLCR time series for the determination of the existing periodicities. The dashed red line is the 95\% confidence level line for the statistical significant test.}
    \label{fig:example_figure}
\end{figure}

\begin{figure}
    \includegraphics[width=\columnwidth]{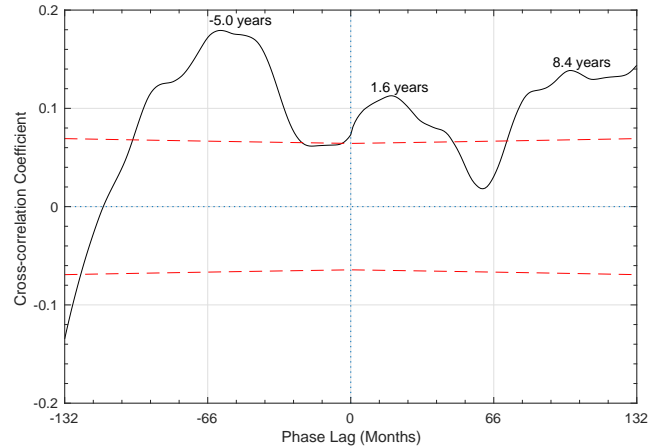}
    \caption{The cross-correlation coefficient between the monthly PLCR and the monthly MCI. The dashed red lines are the 95\% confidence level lines for the statistical significant test.}
    \label{fig:example_figure}
\end{figure}

The periodicity of 6.13 years in the PLCR time series is close to the periodicity of 6.1 years in the differential rotation parameters of Greenwich sunspot groups  \citep{1995SoPh..158..173J}. In the statistical study of \cite{2016SoPh..291.3485J}, a periodicity around 5.4 years was also revealed in the equatorial rotation rate of sunspot groups. They thought that the periodicity of 5.4 years should  be the 3.5 times of the synodic period (1.597 year) of Venus and Earth \citep{2013PRP.....1..147W}. Most recently. by studying the temporal variation of coronal rotation in the 10.7 cm radio flux, \cite{2017ApJ...841...42X} found a periodicity of 6.6 years, which was interpreted as the third harmonic of the 22-year magnetic cycle. A similar periodicity of 6.8 years was found by \cite{2018ApJ...855...84X} who studied the equatorial rotation rate of solar magnetic fields from 1975 September to 2008 April. If the periodicity around 6 years is indeed the 22-year magnetic cycle, then our result is in good agreement with the results given by \cite{2000JApA...21..167J}, \cite{2011MNRAS.414.3158C}, and \cite{2017MNRAS.466.4535B}. That is, the rotational behavior of the solar corona might be related to the 22-year Hale cycle or the magnetic activity reversal. Further studies are needed to reveal the physical origin of the periodicity around 6 years.

The existence of the quasi-eleven-year periodicity in the temporal variation of solar rotation is an open question. According to the statistical analyses of solar radio flux at 2.8 GHz, \cite{2011MNRAS.414.3158C} found that the temporal variation of rotation period does not have any systematic periodicity around 11 years, and there is either not or weak correlation between coronal rotation and solar activity. Later, \cite{2012MNRAS.423.3584L} also studied the cycle-related variation of coronal rotation period, and they claimed that there is no 11-year periodicity for the secular variation of rotation period length. However, \cite{2017ApJ...841...42X} restudied the temporal variation of coronal rotation based on the continuous wavelet transform and auto-correlation analysis, they thought that the long-term variation of coronal rotation period should be related to the 11-year Schwabe cycle, because a periodicity of 10.3 years is detected from the smoothed coronal rotation period. By studying the rotation characteristics of historical solar observations, including the Greenwich sunspot groups, the spectroscopic velocity data, the large-scale magnetic fields, and the photospheric magnetic maps, the quasi-eleven-year periodicity of solar rotation variation has also been reported by \cite{1997SoPh..170..389J}, \cite{1999SoPh..184...41J}, \cite{2001SoPh..201....1O}, \cite{2006SoPh..237..365B}, and \cite{2018ApJ...855...84X}. In this work, we find that the quasi-11-year periodicity (here are 9.53 and 11.13 years) indeed exists in the PLCR time series. So from this, our analysis results strongly support the standpoint that the temporal variation of coronal rotation has a close connection with the 11-year Schwabe cycle.

\subsection{Phase Relationship between Coronal Rotation and Solar Activity}

Earlier studies \citep[see][]{2001SSRv...95..227R,2002SoPh..206..401M,2005AdSpR..35..410M,2007SoPh..241..269M,2014SSRv..186..105E} have shown that the intensity of the green corona is highly correlated with other solar activity indicators as all of them are intrinsically inter-linked through solar magnetism. So, the MCI is applied here as an indicator to describe the temporal variability of solar magnetic activity. To better understand the phase relationship between coronal rotation and solar activity, the cross-correlation analysis between the monthly PLCR and the monthly MCI is performed.  Here, we used the monthly data set to remove many small peaks existing in the daily data set. Figure 11 displays the resulting cross-correlation coefficients varying with the relative phase lags (in months), in which the abscissa is the phase shift of the PLCR versus the MCI with forward shifts given positive values. 

As Figure 11 shows, three significant peaks, above the95\% confidence level, within the time shifts of $\pm$132 months ($\pm$ 11 years) are clearly seen. When the time shifts are -60 months (-5.0 years), 19 months (1.6 years), and 101 months (8.4 years), the values of the cross-correlation coefficient reaches a local maximum. Therefore, the phase relationship between the coronal rotation and the solar activity is very complex, without any systematic pattern.

It is well known that the wavelet transformation analysis is a statistical tool widely used to study the nonlinear and non-stationary signals in the time-frequency space. One of its extension, the cross-wavelet transform (CWT), can be applied to determine the relative phase difference between the two time series \citep{2004NPGeo..11..561G}. This technique has been successful used in a broad range of research aspects of solar physics fields, such as \cite{2015AJ....150..171X}, \cite{2017ApJ...850..120R}, \cite{2017ApJ...851..141X}, \cite{2018JGRA..123.6148G}, \cite{2018ApJ...857...28K}, and so on.

\begin{figure*}
    \includegraphics[width=2.0\columnwidth]{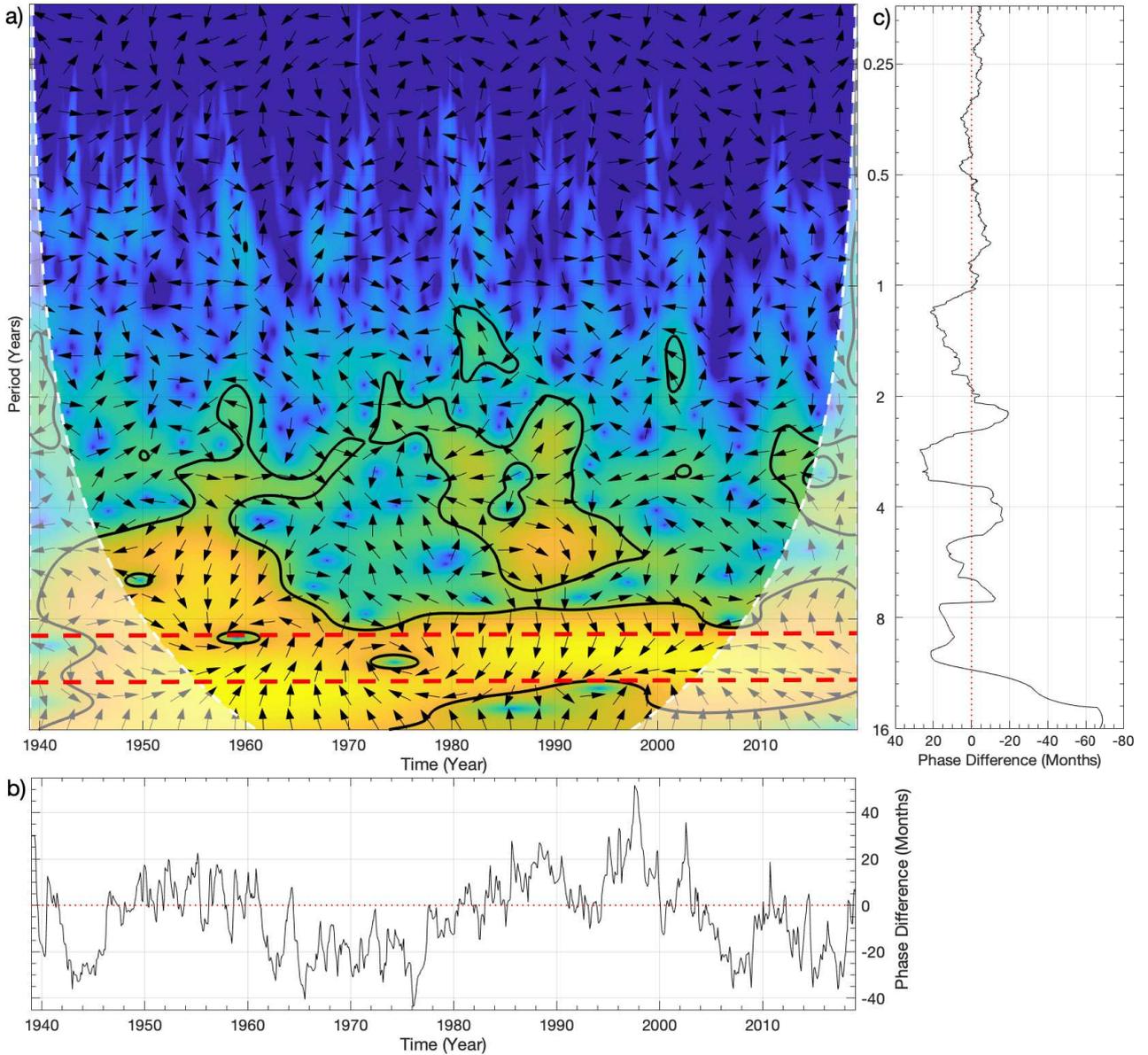}
    \caption{Panel a): the CWT spectra of the monthly PLCR with the monthly MCI, and the dashed white line is the cone of influence (COI) where the wavelet transform suffers from the edges effects. The relative phase relationship is shown as the arrows with in-phase pointing right, anti-phase pointing left, and the PLCR leading the MCI by 90$^\circ$ pointing down. Panel b): the relative phase difference (in
months) between the monthly PLCR and the monthly MCI varying with the time. Panel c): the relative phase differences (in months) between the monthly PLCR and the monthly MCI as a function of the period scales.}
    \label{fig:example_figure}
\end{figure*}

In this work, the codes provided by \cite{2004NPGeo..11..561G}, available in GitHub\footnote{https://github.com/grinsted/wavelet-coherence}, are employed to examine the phase differences varies with the time and the period scale between the monthly PLCR with the monthly MCI. The CWT spectra between them are shown in the panel a) of Figure 12, in which the dashed white line showing the cone of influence (COI) where the local spectrum suffers from the edges effects. In this figure, the relative phase difference is described as the arrows with in-phase pointing right, anti-phase pointing left, and the PLCR lagging behind the MCI by 90$^\circ$ pointing up. From this figure, one can easily see that in the whole frequency ranges (both the high-frequency and the low-frequency components), the power spectrum displays the strong phase mixing, that is, the distribution of the arrow directions (i.e. phase angles) is almost random, implying no regular fluctuation pattern between the two time series.

Based on the local phase relations displayed in the panel a) of Figure 12, we calculated separately the relative phase differences as a function of the time point and the period scale. The results are shown in the panel b) and the panel c) of Figure 12, respectively. Here, the phase difference at a certain time point is simply calculated by averaging all phase differences at that time point in the whole periodical scales. This concept is also used to calculate the relative phase differences varying with the periodic scales. As the two panels display,  the phase relationship between the coronal rotation period and the solar magnetic activity is not only time-dependent but also frequency-dependent.

As shown in Figure 12, when the signal is weak (mostly at high-frequency components), the distribution of the phase is random. If the signal is strong (around the 11-year cycle band), there is a systematic trend in the phase. In the panel a) of Figure 12, the ranges of the 11-year cycle component are indicated as the dashed bold red lines. It is clearly seen that most of the arrows (expect for the time interval 1960-1972) point down, indicating that the PLCR time series leads the MCI time series. According to our results, we should be careful in choosing the suitable periodic ranges and the time interval when we need to study their phase relationship in detail. A similar conclusion has been obtained by \cite{2007A&A...475L..33D} who studied the scale-resolved phase coherence of hemispheric sunspot areas during the time interval from 1874 to 2007. They found that the coherent phase variables of hemispheric sunspot activity are found to exist for a small frequency band with periodicities around the dominating 11-year cycle. So from our analysis, it is reasonably inferred that the small mismatch in the dominant 11-year cycle band brings out the phase to drift.

\section{Conclusions and Discussions}

In the present study, we investigated the long-term temporal variations of solar coronal rotation with the use of the MCI during the time interval between 1939 January 1 and 2019 May 31. Firstly, the continuous wavelet transform was applied to calculate the coronal rotation period from a global point of view. Based on the conception proposed by earlier studies, the time series of the PLCR was extracted. Then, the secular decreasing trend and the possible periodicities of the PLCR time series are obtained by the linear regression analysis and the auto-correlation analysis, respectively. And finally, the cross-correlation analysis and the cross-wavelet transform were used to examine the complex phase relationship between the monthly PLCR and the monthly MCI. 

From a global point of view, the synodic coronal rotation period obtained by us is good agreement with the results reported by previous researchers, whose studies focused on the coronal green line emission, such as \cite{1989ApJ...336..454S}, \cite{1994SoPh..152..161R}, and \cite{1999SSRv...87..211I}. However, the obtained value is slightly greater than the result given by \cite{2011MNRAS.414.3158C} and \cite{2012MNRAS.423.3584L} who studied the periodicities of the coronal rotation derived from the daily radio flux at 2800 MHz. The slightly difference of the rotation period length between the Fe \tiny{XIV} \normalsize{green} line emission and the solar radio flux at 2800 MHz can be interpreted as the following reason. The 10.7 cm solar radio flux is directly related to the total amount of the magnetic flux, which should be originated from the lower (inner) corona at the 60 000 km above the surface atmosphere \citep{2001ApJ...548L..87V,2017SoPh..292...55B}. However, the Fe \tiny{XIV} \normalsize{green} line emission arises in coronal material with a characteristic temperature of about 1.8 $\times 10^{6}$ K \citep{1969MNRAS.142..501J}, which is the typical temperature of most of the corona outside of the coronal holes. Besides, this temperature is also found over active regions, where the Fe \tiny{XIV} \normalsize{emission} reaches its highest intensity. That is, the Fe \tiny{XIV} \normalsize{green} line emission is considered to be related to both the active regions and the quiet corona \citep{1989ApJ...336..454S,2018AstL...44..727B}.

Previous studies \citep[see][]{2009A&A...497..835H,2011Ap&SS.331..441L,2012RAA....12..187X,2017ApJ...841...42X} on the secular trend of the solar rotation period have often revealed a general decreasing trend in the solar rotation period. It should be pointed out that these studies focused on the sunspot activity and the corona radio flux. Our analysis results, based on the database of the Fe \tiny{XIV} \normalsize{green} line emission, somewhat confirm and enhance the earlier findings. So from this work and earlier studies, we infer that the sunspot activity, the coronal radio flux, and the green corona exhibit the similar decrease trend in the rotation period over the past several decades, i.e., the Sun accelerates its global rotation rate in the long run. However, an opposite long-term trend (namely, a secular deceleration) for the solar rotation rate was also obtained by some authors. For example, \cite{2005SoPh..232...25J,2005ApJ...626..579J}, \cite{2004HvaOB..28...55B}, and \cite{2006SoPh..237..365B} found that the solar rotation rate exhibits a secular deceleration trend. The reason for different opinions might be possibly related to the different data analysis methods those are used. The time-frequency signal analysis is applied to estimate the solar rotation period by \cite{2009A&A...497..835H}, \cite{2011ApJ...730...49L}, \cite{2017ApJ...841...42X}, and this work, which is mainly caused by the solar large-scale active regions or active longitudes. However, the solar rotation rate studied by \cite{2005SoPh..232...25J,2005ApJ...626..579J}, and \cite{2004HvaOB..28...55B} is estimated from the position variation of the surface tracers (sunspot or sunspot group). The key problem of interpreting the analysis results derived from different tracers is whether the observed changes represent the global variation of the rotation or might be caused by the specific property of the used tracer, as previously pointed out by \cite{2006SoPh..237..365B}. Further studies are needed to reveal this problem in the future.

Four significant periodicities with the values of 3.25, 6.13, 9.53, and 11.13 years exist in the temporal variation of the coronal rotation period. The first one might be related to the solar QBO, which should be caused by the magnetic Rossby wave instabilities in the solar tachocline \citep{2010ApJ...724L..95Z}. The possible physical origin of the second periodicity (6.13 years) has not fully understood, but it can be interpreted as the third harmonic of the 22-year magnetic cycle. Our analysis result supports the standpoint that the rotational behavior of the solar corona might be related to the 22-year Hale cycle or the magnetic activity reversal, which has been found by \cite{2005ApJ...626..579J}, \cite{2011MNRAS.414.3158C}, and \cite{2017MNRAS.466.4535B}. During the past several years, the existence of the quasi-eleven-year periodicity in the temporal variation of solar rotation is an open question, and many authors have given differing opinions. Based on our analyses, the last two periodicities (9.53 years and 11.13 years) can be considered as the quasi-11-year Schwabe cycle. That is, the temporal variation of the solar coronal rotation has a close connection with the eleven-year Schwabe cycle.
 
The relative phase relationship between the PLCR and the MCI, obtained by the cross-correlation analysis, does not display any regular pattern. That is, their phase relationship is very complex, which is also confirmed and enhanced by the CWT analysis. For both the high-frequency and the low-frequency components, the power spectrum shows the strong phase mixing and the distribution of the arrow directions is almost random. By calculating the relative phase differences as a function of the time point and the period scale, we found that the phase relationship between the coronal rotation period and the solar magnetic activity is not only time-dependent but also frequency-dependent. It should be noticed that the suitable periodic component and the time interval should be carefully chose and considered. For a small range around the 11-year cycle band, there is a systematic trend in the phase, and the small mismatch in this band brings out the phase to drift. To better understand the relationship between the coronal rotation and the solar magnetic activity, it will be very interesting to further investigate their complicated phase relationship in the future.

\section*{Acknowledgements}

The database  of daily and monthly modified coronal index used in this study was introduced by B. Luk\' a\u c and M. Rybansk\' y, and can be downloaded from the website of the Slovak Central Observatory in Hurbanovo (http://www.suh.sk/obs/vysl/MCI.htm). The sunspot number data was obtained from the WDC-SILSO, Royal Observatory of Belgium, Brussels (http://www.sidc.be/silso/). This work is supported by the National Key Research and Development Program of China (2018YFA0404603), the Joint Research Fund in Astronomy (Nos. U1631129, U1831204, U1931141) under cooperative agreement between the National Natural Science Foundation of China (NSFC) and Chinese Academy of Sciences (CAS), the National Natural Science Foundation of China (Nos. 11873089, 11903009), the Youth Innovation Promotion Association CAS, the Yunnan Key Research and Development Program (2018IA054), the open research program of CAS Key Laboratory of Solar Activity (Nos. KLSA201807), the Key Laboratory of Geospace Environment, CAS, University of Science \& Technology of China, the major scientific research project of Guangdong regular institutions of higher learning (2017KZDXM062), and the youth project of science and technology research program of Chongqing Education Commission of China (KJQN201801325). Last, but not least, the authors wish to express their gratitude to the anonymous referee whose supporting recommendations provided much help in improving the original manuscript.

\bibliographystyle{mnras}
\bibliography{mnras} 

\bsp	
\label{lastpage}
\end{document}